\documentclass[
    ,final            
  ]
  {aipproc}

\layoutstyle{6x9}

\usepackage{amssymb,latexsym,amsmath}
\begin{document}

\title{High-Frequency QPOs as a Product of Inner Disk Dynamics around Neutron Stars}


\author{M. Hakan Erkut}{
  address={Physics Department, \.Istanbul K\"{u}lt\"{u}r
University, Atak\"{o}y Campus, Bak\i rk\"{o}y 34156, \.Istanbul,
Turkey} }

\begin{abstract}
The kHz QPOs observed in a neutron star low mass X-ray binary are
likely to be produced in the innermost regions of accretion disk
around the neutron star. The rotational dynamics of the inner disk
can be characterized by the presence of either sub-Keplerian or
super-Keplerian accretion flow depending on the relative fastness of
the neutron spin as compared to the Keplerian frequency at the inner
disk radius. Within the magnetosphere-disk interaction model, the
frequency difference between the two kHz QPOs observed in a given
source can be estimated to be slightly higher than or nearly around
the neutron star spin frequency if the neutron star is a slow
rotator and less than the stellar spin frequency if the neutron star
is a fast rotator.
\end{abstract}

\maketitle


\section{Introduction}

The observations of neutron stars in low mass X-ray binaries (LMXBs)
led to the discovery of the kHz quasi-periodic oscillations (QPOs)
and the burst oscillations \cite{vdK00}. It has almost been firmly
established from the observations that the frequency of the burst
oscillation is very close to the spin frequency of the neutron star,
i.e., $\nu _{\mathrm{burst}}\simeq \nu _{\mathrm{spin}}$
\cite{MB07}.

The kHz QPOs from neutron star LMXBs have some important
observational properties. These high-frequency QPOs usually appear
in pairs in the $\simeq 200-1200$ Hz range. Both the upper kHz QPO
frequency $\nu _{2}$ and the lower kHz QPO frequency $\nu _{1}$
correlate with the X-ray luminosity of the source \cite{vdK00}.
According to early observations, the peak separation between the kHz
QPO frequencies seemed to be directly related to the spin frequency
of the neutron star, i.e., $\Delta \nu =\nu _{2}-\nu _{1}\simeq \nu
_{\mathrm{burst}}\simeq \nu _{\mathrm{spin}}$. The systematic
analysis of all QPO data, however, showed that the peak separation
$\Delta \nu $ is not constant for a given source. Indeed, $\Delta
\nu $ decreases as both the kHz QPO frequencies $\nu _{2}$ and $\nu
_{1}$ increase \cite{Petal98}. According to the analysis of
the early kHz QPO data, $\Delta \nu \simeq \nu _{\mathrm{burst}}\simeq \nu _{\mathrm{spin}%
} $ for $\nu _{\mathrm{spin}}<400$ Hz and $\Delta \nu \simeq \nu _{\mathrm{%
burst}}/2\simeq \nu _{\mathrm{spin}}/2$ for $\nu
_{\mathrm{spin}}\gtrsim 400$ Hz among neutron star sources with
different burst or spin frequencies \cite{vdK06}. The recent
analysis of all available data, however, revealed that $\Delta \nu $
significantly deviates from either $\nu _{\mathrm{spin}}$ or $\nu _{\mathrm{%
spin}}/2$ for almost half of all sources and that $\Delta \nu /\nu _{\mathrm{%
spin}}$ decreases from a value $>1$ to a value $<1/2$ as $\nu _{\mathrm{spin}%
}$ increases \cite{MB07}.

The frequency correlations of kHz QPOs can be explained, albeit
qualitatively at this stage, within the boundary region model
\cite{AP08,EPA08}. Next, we discuss the interpretation of the
observational properties of kHz QPOs within the boundary region
model.

\section{Boundary region model}

According to the boundary region model (BRM), the likely origin of
the high-frequency QPOs is the innermost region of the accretion
disk around the compact object. In neutron star sources, the kHz
QPOs are produced in the boundary region (BR) or the boundary layer
where the angular velocity of the accretion flow is non-Keplerian
\cite{AP08,EPA08}. In black hole candidates, it is the relativistic
disk region near the innermost stable circular orbit which is
responsible for the production of the high-frequency QPO pairs with
a frequency ratio around 1.5 \cite{E10}.

The interaction of the inner disk with the magnetosphere of the
neutron star rotating with an angular frequency $\Omega
_{\mathrm{spin}}=\Omega _{\ast }$ leads to the formation of a
magnetohydrodynamic BR throughout which the angular frequency
$\Omega $ of the accreting matter deviates from the Keplerian
frequency $\Omega _{\mathrm{K}}$ to match the stellar spin frequency
$\Omega _{\ast }$ at the innermost disk radius $r_{\mathrm{in}}$
\cite{EA04}.

In the conventional regime of accretion where the neutron star is a \emph{%
slow rotator}, $\Omega _{\ast }<\Omega
_{\mathrm{K}}(r_{\mathrm{in}})$ and the angular frequency profile of
the matter is characterized by a maximum frequency $\Omega _{\max
}=\Omega (r_{0})$ at some radius $r_{0}$ in the BR \cite{EA04}. In
the Newtonian regime, the radial epicyclic
frequency $\kappa $ is the same as the angular frequency $\Omega $ if $%
\Omega =\Omega _{\mathrm{K}}$. It is a well known fact that the
nondegeneracy between $\kappa $ and $\Omega $ appears in the regime
of strong gravity due to general relativistic effects. Even in the
non-relativistic regime, the degeneracy between $\kappa $ and
$\Omega $ is removed in a hydrodynamic BR throughout which $\kappa
>\Omega $ if the BR is sub-Keplerian \cite{AP08,EPA08}. The analysis of global hydrodynamic modes
of free oscillations in a typical sub-Keplerian BR revealed that the
fastest growing modes are excited near the disk radius $r_{0}$ where
$\Omega $ is maximum. The frequencies of the growing hydrodynamic
modes match the test-particle frequency branches $\kappa \pm \Omega
$ and $\kappa $ in the limit of small hydrodynamic corrections. The
same modes grow in the $\simeq r_{\mathrm{in}}-r_{0}$ range. The
difference between the two consecutive frequency bands of the
growing modes is $\sim \Omega $ which corresponds to the frequency
separations $\Delta \omega \simeq \Omega _{\max }\simeq
1.2-1.3\Omega _{\ast }$ and $\Delta \omega \simeq \Omega _{\ast }$
at $r=r_{0}$ and $r=r_{\mathrm{in}}$, respectively \cite{EPA08}.
Note that the estimations of the BRM for the peak separation of the
growing mode frequencies are in agreement with the distribution of
the $\Delta \nu /\nu _{\mathrm{spin}}$ values observed
for the relatively slowly rotating neutron stars with spin frequencies below $%
\sim 400$ Hz \cite{MB07}. The frequencies of the modes which grow in
amplitude at any particular radius in the BR increase all together
as the local sound speed $c_{s}$ increases \cite{EPA08}. The
magnetic field strengths of neutron stars in LMXBs are generally
thought to be weak enough for the radiation pressure to be important
in the innermost disk regions. For a radiation pressure dominated
inner disk, $c_{s}\propto \dot{M}$, where $\dot{M}$ is the mass
inflow rate in the inner disk \cite{SS73}. We expect, in accordance
with the correlations of the kHz QPO frequencies, that the growing
mode frequencies correlate with the X-ray luminosity
$L_{\mathrm{X}}\propto \dot{M}$. Figure~$1$ exhibits the correlation
between the frequencies $\nu _{1}$ and $\nu _{2}$
of two consecutive modes growing at the innermost disk radius $r_{\mathrm{in}%
}$ and the correlation between the frequency difference $\Delta \nu
=\nu _{2}-\nu _{1}$ and the upper mode frequency $\nu _{2}$ of the
same modes estimated by the BRM \cite{EPA08} for a putative neutron
star with a spin frequency $300$ Hz. Note from Figure~$1$ that the
power-law fit to the kHz QPO data of Sco X-1 \cite{Petal98} can also
be used to fit the sample data estimated by the BRM simply by
adjusting the proportionality constant in the power law.

\begin{figure}[htbp]
  \centering
  \includegraphics[width=0.5\textwidth]{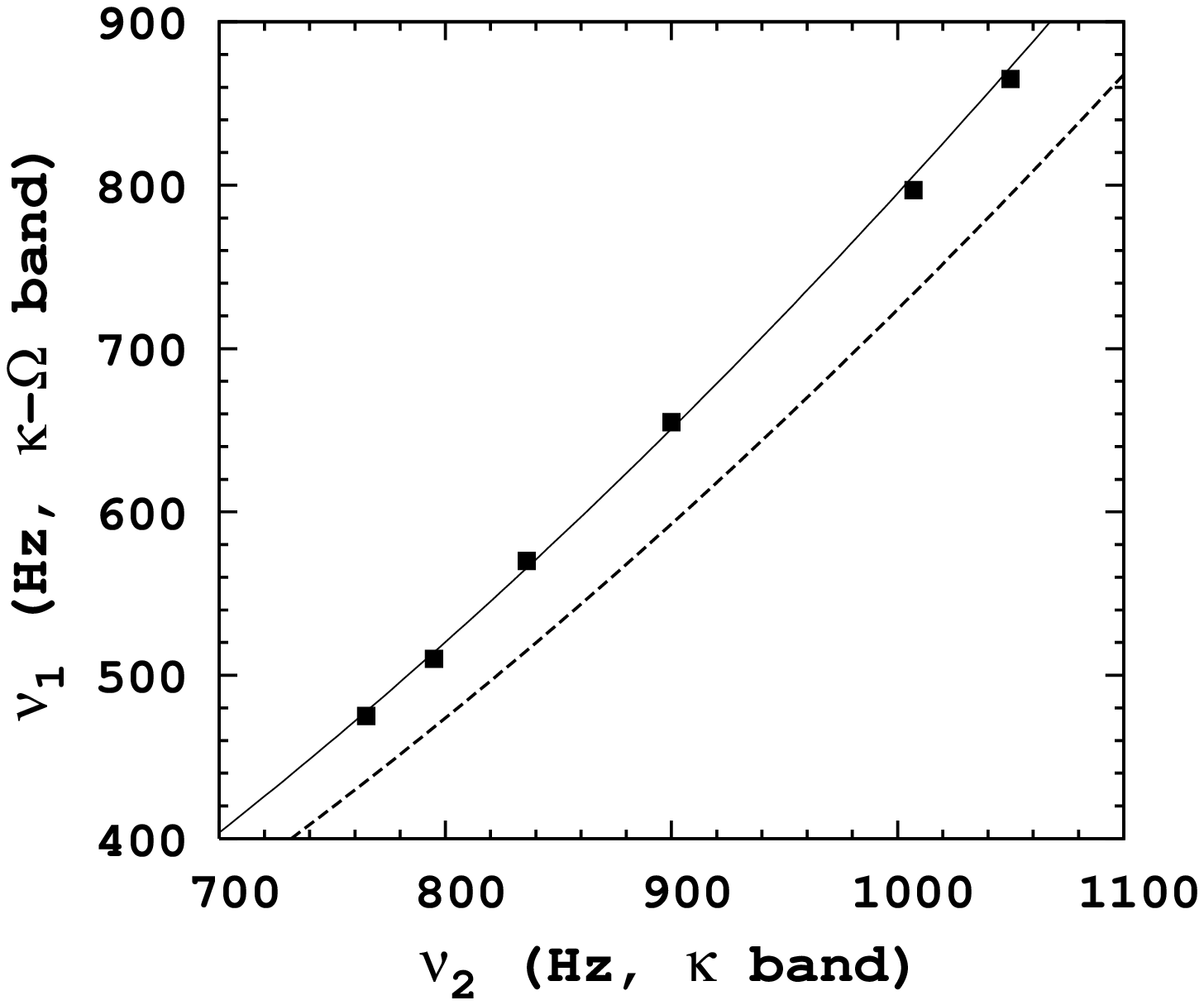}
  \includegraphics[width=0.5\textwidth]{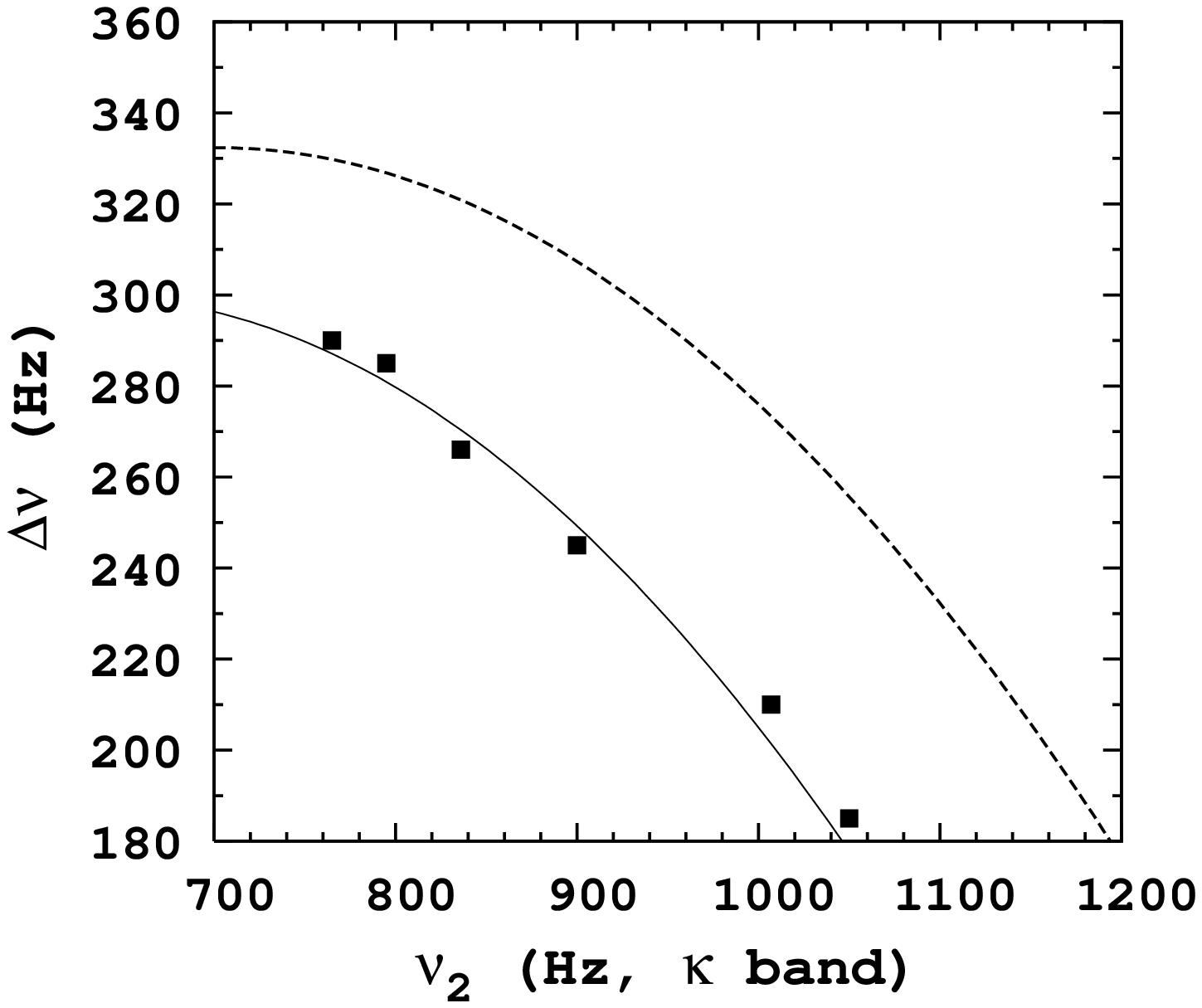}
  \caption{Power-law fits to the correlation of the frequencies
  $\nu _{1}$ and $\nu _{2}$ and to the correlation between the frequency
  difference $\Delta \nu$ and $\nu _{2}$ estimated by the BRM for the
  growing modes at the innermost disk radius. The power law is given by
  $\nu _{1}=C(\nu _{2}/1000)^{1.9}$ Hz, where $C=795$ for the solid
  curve and $C=724$ for the dashed curve \cite{Petal98}.}
  \label{fig:1}
\end{figure}

According to the present distribution of $\Delta \nu /\nu
_{\mathrm{spin}}$ over $\nu _{\mathrm{spin}}$ \cite{MB07}, the
relatively fast rotating neutron stars with
$\nu_{\mathrm{spin}}\gtrsim 400$ Hz seem to exhibit kHz QPO pairs
for which $\Delta \nu $ is around or sometimes even less than $\nu
_{\mathrm{spin}}/2$. The systematic trend of observing small values
of $\Delta \nu /\nu _{\mathrm{spin}}$ for sufficiently high values
of $\nu _{\mathrm{spin}}$ could be the result of a change in the
structure of the BR. In the magnetosphere-disk interaction model,
the structure of the BR depends on the fastness parameter, $\omega
_{\ast }=\Omega _{\ast }/\Omega _{\mathrm{K}}(r_{\mathrm{in}})$. The
neutron star is a \emph{slow rotator} if $\omega _{\ast }<1$. For a
\emph{slow rotator}, the BR is sub-Keplerian as mentioned above. In
the accretion regime where the neutron star is a \emph{fast rotator}
($\omega _{\ast }\geq 1$), we expect to find a super-Keplerian BR for $r_{\mathrm{in}}\leq r\leq r_{%
\mathrm{A}}$, where $r_{\mathrm{A}}$ is the effective magnetic
coupling radius, e.g., the Alfv\'{e}n radius. The magnetic coupling radius, for a \emph{%
fast rotator}, exceeds the corotation radius $r_{\mathrm{co}}$, where $%
\Omega _{\ast }=\Omega _{\mathrm{K}}$. The angular frequency $\Omega
$ of the disk matter accreting through the super-Keplerian BR
deviates from the Keplerian frequency $\Omega _{\mathrm{K}}$ to
match the stellar spin
frequency $\Omega _{\ast }$ at $r=r_{\mathrm{in}}$. Condition for the \emph{%
fast rotator} to be an accreting neutron star and not to be a
propeller can
be written as $\Omega _{\mathrm{K}}\leq \Omega \leq \sqrt{2}\Omega _{\mathrm{%
K}}$ for $r_{\mathrm{in}}\leq r\leq r_{\mathrm{A}}$. Thus, a
\emph{fast rotator} accretes only within a limited range of the
fastness, i.e, $1\leq \omega _{\ast }\leq \sqrt{2}$. In Figure~$2$,
we display the run of the dynamical frequencies $\kappa $ (dashed
curve) and $\Omega $ (dotted curve) throughout the super-Keplerian
BR for a \emph{fast rotator} with $\omega_{\ast }=1.4$. The angular
frequency $\Omega $ lies in the $\Omega _{\mathrm{K}}-\sqrt{2}\Omega
_{\mathrm{K}}$ range (region bounded by the solid cuves) as
expected. The radial epicyclic frequency $\kappa $ is always less
than $\Omega _{\ast }$ and sometimes even less than
$\Omega_{\ast}/2$. As compared to the right panel of Figure~$2$,
$r_{\mathrm{A}}$ has a greater value in the left panel. Note that
$\kappa (r_{\mathrm{in}})$ decreases as $r_{\mathrm{A}}$ decreases.
Although the mode analysis is necessary to distinguish among
different hydrodynamic modes, it is noteworthy to find the frequency
separation of two consecutive modes to be $\Delta \omega \simeq
\kappa $ if the high-frequency modes with $\Omega \pm \kappa $ and
$\Omega $ branches determine the kHz QPOs in the \emph{fast rotator}
regime. For the \emph{fast rotators}, $\Delta \omega $ decreases as
$\dot{M}$ increases since $r_{\mathrm{A}}\propto \dot{M}^{-2/7}$ for the Alfv%
\'{e}n radius.

\begin{figure}[htbp]
  \centering
  \includegraphics[width=0.5\textwidth]{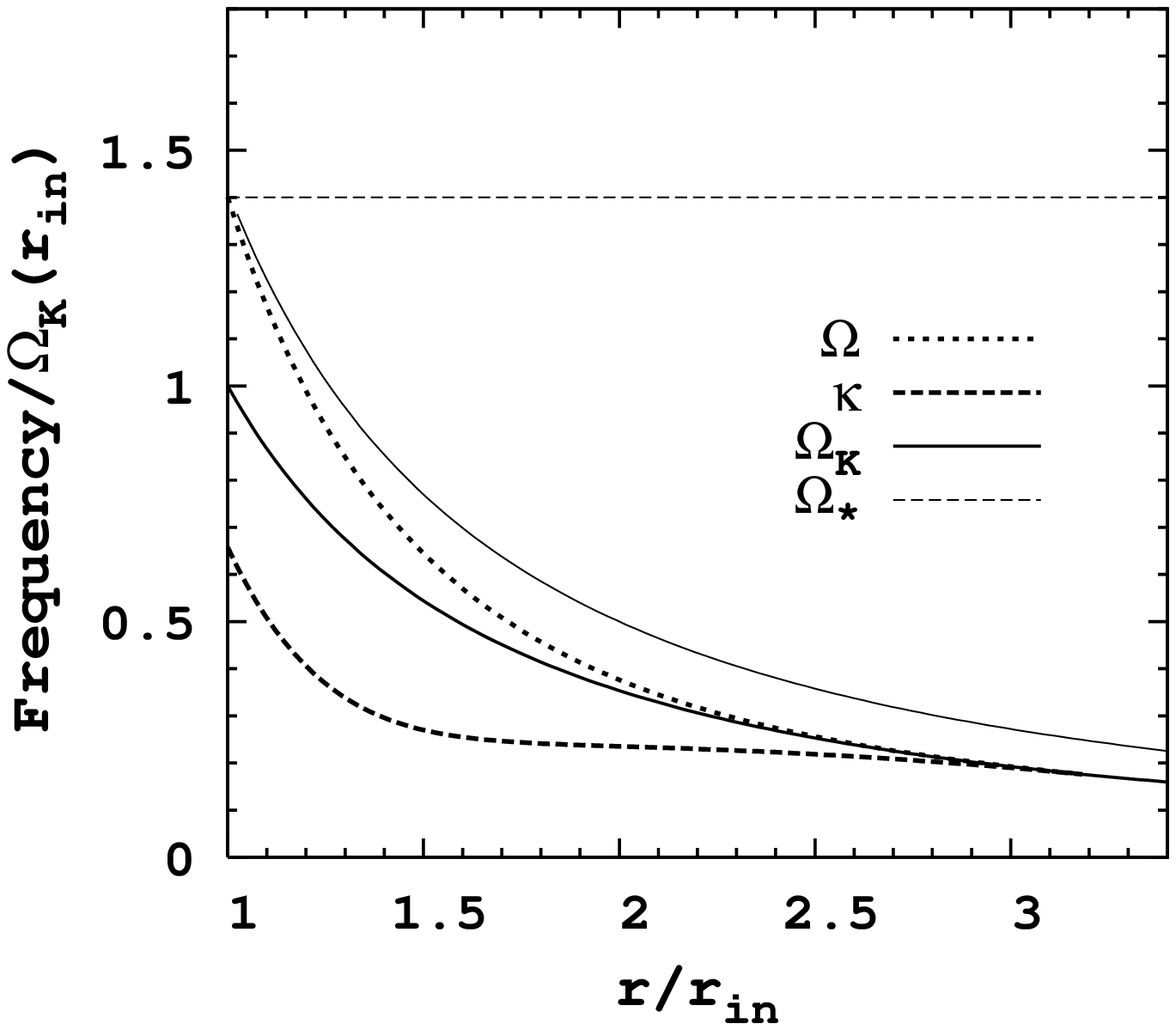}
  \includegraphics[width=0.5\textwidth]{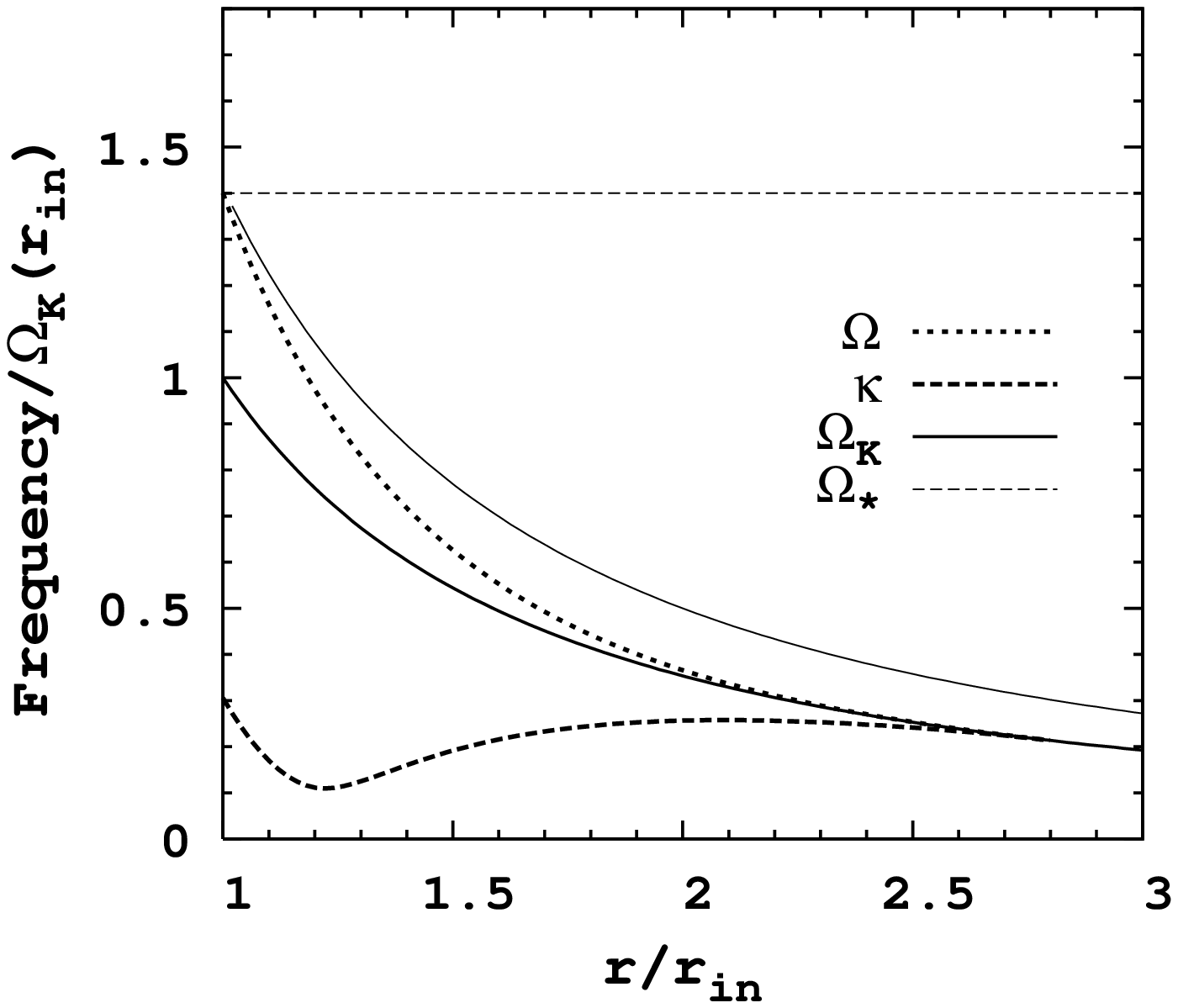}
  \caption{The radial profiles of the dynamical frequencies $\kappa $
  (dashed curve) and $\Omega $ (dotted curve) in the BR of the accretion
  disk around a rapidly rotating neutron star with a fastness $\omega_{\ast }=1.4$.
  The two solid curves correspond to the frequencies $\Omega _{\mathrm{K}}$
  and $\sqrt{2}\Omega _{\mathrm{K}}$.}
  \label{fig:2}
\end{figure}

\section{Conclusions}

The BRM estimations for the peak separations and correlations of kHz
QPOs are in agreement with observations. A sub-Keplerian BR is
appropriate for the \emph{slow rotators} that exhibit kHz QPOs with
the peak separations that are slightly higher than or nearly around
or slightly less than the neutron star spin frequency. Within the
magnetosphere-disk interaction model, super-Keplerian BRs are more
appropriate for the accretion disks around \emph{fast rotators}. In
a super-Keplerian BR, the frequency difference of the mode branches
is always less than the stellar spin frequency if the neutron star
is a \emph{fast rotator}.


\begin{theacknowledgments}
  I would like to thank M. A. Alpar and D. Psaltis for useful
  discussions.
\end{theacknowledgments}

\bibliographystyle{aipproc}   


\end{document}